# Excitonic nonlinear absorption in CdS nanocrystals studied using Z-scan technique

J. He, W. Ji,[a)] G. H. Ma, S. H. Tang, H. I. Elim, and W. X. Sun
Department of Physics, National University of Singapore,
2 Science Drive 3, Singapore 117542, Singapore

Z. H. Zhang and W. S. Chin[a)]
Department of Chemistry, National University of Singapore,
2 Science Drive 3, Singapore 117542, Singapore

a) Author to whom correspondence should be addressed; Electronic email: phyjiwei@nus.edu.sg

**Abstract**

Irradiance dependence of excitonic nonlinear absorption in Cadmium Sulfide (CdS) nanocrystals has been studied by using Z-scan method with nanosecond laser pulses. The wavelength dependence of nonlinear absorption has also been measured near the excitonic transition of 1S(e) – $1S_{3/2}$(h). We observe the saturable absorption, which can be described by a third-order and a fifth-order nonlinear process for both 3.0-nm-sized and 2.3-nm-sized CdS nanocrystals. The experimental results show that the excitonic nonlinear absorption of CdS nanocrystals is greatly enhanced with decreasing particle size. A two-level model is utilized to explain both irradiance and wavelength dependence of the excitonic nonlinearity.

## I. INTRODUCTION

In the past decade, there has been increasing interest in the nonlinear optical and luminescent properties of nanometer-sized semiconductor crystals.[1-9] This interest is stimulated by remarkably different optical properties when semiconductor sizes are decreased from bulk to a few nanometers as predicted by quantum confinement theory. Such unusual properties may have technological applications such as optical-switching devices, quantum-dot lasers, or light emitters. It is well-known that the linear optical properties of semiconductor nanoparticles depend strongly upon particle size, for example, the blue shift of excitonic absorption and emission peaks with decreasing particle size. Additionally, large optical nonlinearities in CdS nanocrystals (NCs) have been observed by degenerate four-wave mixing (DFWM) and pump-probe measurements with nanosecond and picosecond laser pulses.[10-14] And the excited state dynamics have been studied in detail with femtosecond time-resolved pump-probe and photoluminescence (PL) technique.[3, 15, 16] Compared to DFWM and pump-probe technique, Z-scan is an effective technique to investigate nonlinear absorption and nonlinear refraction separately. Recently, there have been several reports on CdS nanoparticles studied by the use of Z-scan technique.[4, 5, 8] In this paper, we report an investigation to the irradiance dependence of pure nonlinear absorption in polymer-stabilized CdS NCs. The investigation has been carried out with nanosecond Z-scan method at photon energies near the lowest exciton energy. The excitonic nonlinear absorption in CdS NCs is observed to increase with decreasing particle size, demonstrating that the nonlinear optical properties of CdS NCs in the strong quantum confinement region are greatly enhanced compared to their bulk counterpart. A two-level model is utilized to explain the irradiance and wavelength dependence of the observed absorption saturation near the lowest exciton peak. The agreement between the model and experimental results reveals the oscillator strengths for CdS NCs of different sizes. The size dependence of the oscillator strength confirms again the quantum confinement effect. The saturation intensity near the lowest exciton resonance is also determined.

## II. EXPERIMENTAL DETAILS

Two samples with CdS NCs of different sizes have been investigated in the present work. In Sample 1 and Sample 2, the CdS NCs capped with dodecanethiol, are 3.0 nm and 2.3 nm in diameter, respectively. The sample preparation procedures are briefly described as follows: 10 mMol dodecanethiol was dissolved into 25 mL toluene. Then 54 mg 2,2'-bipyridyl cadmium(II) thiolcarboxylate [(2, 2'-bpy)Cd(SC{O}Ph)$_2$] as the precursor was added into the solution and the mixture was bubbled with $N_2$ for 30 minutes. Subsequently, the solution was heated slowly and refluxed for 30 minutes. After cooling down, the clear yellowish solution was condensed using a rotary evaporator and the particles were precipitated by the addition of acetone or ethanol. The solid was thoroughly washed with ethanol, dimethyl formamide (DMF) and ether. And then it was vacuum-dried. The resulting powder is dissolvable in nonpolar solvent such as toluene, hexane and chloroform. In our case, the 3.0-nm-sized and 2.3-nm-sized CdS NCs were dissolved in chloroform with volume fraction of $0.12 \times 10^{-3}$ (or $4 \times 10^{-3}$ Mol/L) and $0.06 \times 10^{-3}$ (or $2 \times 10^{-3}$ Mol/L), respectively. More details can be found in Ref. 17.

The optical nonlinearities were measured by using standard Z-scan method with full width at half maximum (FWHM) 5-ns laser pulses produced by an optical parametric oscillator (Spectra Physics MOPO 710). The optical parametric oscillation (OPO) laser pulses were delivered at 10-Hz repetition rate, and were focused onto the sample with a minimum beam waist of 53 $\mu$m. The spatial distribution of the pulse was nearly Gaussian after passing through a spatial filter. The pulse was divided by a beam splitter into two parts. The reflected part was taken as the reference representing the incident pulse energy and the transmitted beam was focused through the sample. Two light energy detectors (Laser Precision RjP-735) were used to record the incident and transmitted pulse energies simultaneously. The sample was placed in a 10-mm-thick quartz cell and mounted on a computer-controlled translation stage, facilitating the sample movement along the Z-axis. All the Z-scan measurements were carried out at room temperature. Thermal effects were ignored since the low repetition rate of 10 Hz was used. Pure solvent was also examined at the same experimental conditions and its nonlinear response was found to be insignificant.

The morphology and distribution of CdS NCs were inspected with a transmission electron microscope (TEM, JEOL-JEM 2010F) operating at 200 kV. The CdS NCs were dissolved in toluene and a drop of the solution was placed on a copper grid with carbon film. It was dried in desiccators before transferring into the TEM sample chamber. X-ray diffraction diffractograms (XRD) of the CdS NC powders were recorded on a Siemens D5005 X-ray powder diffractometer with Cu $K\alpha$ radiation (40 kV, 40 mA). The CdS NC powder was mounted on a sample holder and scanned with a step width of 0.01° in the range from 20° to 60°. Optical absorption spectra of the samples were recorded on a UV-visible spectrophotometer (UV-1601, Shimadzu) in the range from 300 to 800 nm. All the spectra were corrected with pristine solution. To check photo-stability of the two samples, we measured the absorption spectra before and after the laser irradiation and no difference was observed.

## III. RESULTS AND DISCUSSION

Figure 1(a) presents a TEM image of CdS NCs capped with dodecanethiol in Sample 2 and the size distribution with the average diameter of 2.3 ± 0.3 nm is shown in Fig. 1(b). The solid line is a lognormal fit to the size distribution.

A typical XRD pattern of the dodecanethiol-capped CdS NCs in Sample 2 is shown in Fig.2. XRD of the CdS NCs reveals diffraction peaks corresponding to the cubic (zinc blend) structural form. Compared to the XRD of bulk CdS, the present diffraction lines are generally broader. The finite size of the crystallites, strain, and instrumental effects contribute to the broadening. The particle size of the quantum dots (QDs) is determined by using the Scherrer formula,[18]

$$D = \frac{0.9\lambda}{\beta_{size}\cos\theta} \tag{1}$$

where $D$ is the mean diameter of the CdS NCs, $\lambda$ the wavelength of the X rays used and $\beta_{\mathrm{size}}$ the FWHM of the XRD peak corresponding to the Bragg angle 2θ. In order to determine the full width at half maximum, the peak profile is obtained by fitting observed diffraction pattern with Gaussian curves. The calculation shows that the CdS NCs in Sample 2 have a mean diameter of 1.7 nm, which is close to the value of 2.3 nm based on the TEM analysis. As for the CdS NCs in Sample 1, the mean diameter of ~ 3.0 nm is obtained by TEM image.[17]

Optical absorption spectra of the dodecanethiol-capped CdS NCs in chloroform (Sample 1 and Sample 2) are shown in Figs. 3(a) and 3(b), respectively. The absorption onsets are localized at 435 and 425 nm for CdS NCs of 3.0-nm and 2.3-nm diameter, respectively. Comparing with the absorption edge of the bulk CdS (~ 520 nm), the absorption edges for the CdS NCs in the two samples show a blue shift of 85 and 95 nm, correspondingly to 0.47 and 0.53 eV, respectively. The Gaussian fitting curves clearly indicate the presence of the lowest excitonic transition, $1S(e) – 1S_{3/2}(h)$, centered at 400 nm for Sample 1 and 380 nm for Sample 2, which confirms the size effect in CdS NCs. The lowest exciton peak is obviously shifted from 400 nm to 380 nm as the size decreases from 3.0 nm to 2.3 nm. The exciton peaks are in close agreement with the experimental findings of Wang *et al.*[19] and a recent theoretical prediction based on approximation of hole effective-mass Hamiltonian and inclusion of Coulomb interaction.[20] In addition, the sharp absorption increase of the exciton peak reveals narrow size distribution. The broadening of optical transitions observed in the absorption spectra of CdS NCs is primarily due to inhomogeneity arising from size dispersion. In the strong confinement regime, the size dependence of energy transitions in NCs is dominated by a term proportional to $1/R^2$. In this case, the inhomogeneous broadening ($\Gamma_i$) of the transition at energy $\hbar\omega_i$ is proportional to the shift of this transition with respect to the bulk material energy gap ($E_g$). For a Gaussian distribution with a standard deviation $\Delta_R$, $\Gamma_i$ is equal to $2(\hbar\omega_i - E_g)\Delta_R/R$. For CdS NCs with $2R = 2.3$ nm, $\hbar\omega_1 - E_g = 840$ meV, and $\Delta_R/R = 13\%$, we calculate $2\Gamma_1 = 440$ meV, which is consistent with the 450-meV width of the excitonic transition of $1S(e) - 1S_{3/2}(h)$ as shown in Fig. 3(b).

In order to systematically investigate the nonlinear absorption, we performed open aperture nanosecond Z-scan with wavelengths close to the lowest exciton energy. Fig. 4(a) shows typical Z-scan curves of Sample 1 with different wavelengths under an input irradiance of 12 $MW/cm^2$. (Herein the input irradiance is defined as the maximum irradiance at the focal position on the Z-axis). The open aperture Z-scans were only performed at wavelengths longer than 420 nm due to the limitation of OPO laser output. Thus, Z-scans were conducted at wavelengths on the lower-energy side of the lowest exciton peak for both Sample 1 and Sample 2. As shown in Fig. 4(a), all Z-scan curves behave saturable absorption (SA) and the SA signals increase as operating wavelength approaching the exciton peak. To fit the Z-scan data, we employ the standard Z-scan theory [21] with assumption of $\alpha = \alpha_0 + (\alpha_2 + \alpha_4 I) I$, where $\alpha$ and $\alpha_0$ are the total and linear absorption coefficient, respectively, $\alpha_2$ is the third-order nonlinear absorption coefficient, $\alpha_4$ denotes the fifth-order nonlinearity and I represents the laser irradiance. Fig. 4(b) shows the comparison of open aperture Z-scan measurements of 3.0-nm-diameter CdS NCs under different irradiances at 430 nm. The symbols are experimental data while the solid curves are theoretically fitting curves by the Z-scan theory including both $\alpha_2$ and $\alpha_4$ nonlinear terms. It is interesting to note that a transmission decrease occurs near the focal position under the higher irradiance, which indicates a higher order nonlinear absorption process. We also employ the standard Z-scan theory with assumption of $\alpha = \alpha_0 + \alpha_{2eff} I$ to fit the experimental Z-scan data, where $\alpha_{2eff} = \alpha_2 + \alpha_4 I$. Fig.4 (c) displays the best-fit values of $\alpha_{2eff}$ at 430 nm as a function of input irradiance for Sample 1. A linear relationship is obtained as expected. Thus, the $\alpha_2$ and $\alpha_4$ values can be extracted from the intercept at the Y-axis and the slope of the linear fit, respectively. The deduced $\alpha_2$ and $\alpha_4$ values quantitatively agree with the previous fitting results with assumption of $\alpha = \alpha_0 +(\alpha_2 + \alpha_4 I) I$. Fig. 5 shows wavelength dependence of open aperture Z-scan measurements of 3.0-nm-diameter CdS NCs at high irradiance (40 MW/cm$^2$). The symbols are experimental data while the solid curves are the theoretically fitting curves by the standard Z-scan theory. At this high irradiance, the onset wavelength for occurrence of transmission decrease is ~ 435 nm, which just corresponds to the absorption bandedge of the CdS NCs. The relative decrease in the transmission at focus varies with wavelength, indicating that the observed dip is due to the fifth-order nonlinear absorption of the CdS NCs. Similar results are obtained for the 2.3-nm-diameter CdS NCs. It is known that the nonlinear absorption coefficients are related to the third-order and fifth-order imaginary susceptibilities by

$$\alpha_2 = 4\pi \,\mathrm{Im}\,\chi^{(3)} \big/ \left(\lambda n_0^{\,2} c \varepsilon_0\right) \tag{2}$$

$$\alpha_4 = 5\pi \,\mathrm{Im}\,\chi^{(5)} \big/ \left(\lambda n_0^{\,3} c^2 \varepsilon_0^{\,2}\right) \tag{3}$$

where $n_0$ is the linear refractive index, $\lambda$ the laser wavelength and c the speed of light in vacuum. Thus, the intrinsic third-order and fifth-order susceptibilities of CdS NCs can be obtained by the use of Eqs. (2) and (3) with the relation of $\mathrm{Im}\chi^{(3)} = f_v |f|^4 \mathrm{Im}\chi^{(3)}_{QD}$ and $\mathrm{Im}\chi^{(5)} = f_v |f|^6 \mathrm{Im}\chi^{(5)}_{QD}$, respectively, where $f_v$ is the volume fraction of CdS NCs in the solution and $f$ is the local field correction that depends on the dielectric constant of solvent and NCs. The value of $f$ is nearly equal to 0.65 in this case. Table I lists the obtained values of $\alpha_2$, $\mathrm{Im}\chi^{(3)}_{QD}$, $\alpha_4$ and $\mathrm{Im}\chi^{(5)}_{QD}$ at different wavelengths for Sample 1 and Sample 2.

To interpret the results quantitatively, a two-level model was employed.[22] By considering the population redistribution under the laser radiation of frequency $\omega$, the susceptibility, $\chi = \mathrm{Re}\chi - i\,\mathrm{Im}\chi$, can be derived as a function of $|E_0|^2$, where $E_0$ is the electric field strength of the laser light. To obtain the third-order and fifth-order nonlinear absorption, $\mathrm{Im}\chi^{(3)}$ and $\mathrm{Im}\chi^{(5)}$ can be expanded in terms of $|E_0|^2$ and written down in the following:

$$\mathrm{Im}\,\chi^{(3)} = -\frac{\mu^4 \mathrm{T}_2{}^2 \tau \Delta \mathrm{N}_0}{\varepsilon_0 \hbar^3} \frac{1}{\left[1 + (\omega - \omega_0)^2 \mathrm{T}_2{}^2\right]^2} \tag{4}$$

$$\mathrm{Im}\,\chi^{(5)} = \frac{\mu^6 \mathrm{T}_2{}^3 \tau^2 \Delta \mathrm{N}_0}{\varepsilon_0 \hbar^5} \frac{1}{\left[1 + (\omega - \omega_0)^2 \mathrm{T}_2{}^2\right]^3} \tag{5}$$

where $\mu$ is the dipole moment, $T_2$ the dephasing time, $\tau$ the relaxation time, $\Delta N_0$ the population difference at zero field, $\varepsilon_0$ the permittivity in vacuum, $\hbar$ the Plank's constant, and $\omega_0$ the exciton frequency. Up to date no accurate calculation can be found in literature for the dipole moment, $\mu = \langle\Phi_1|er|\Phi_2\rangle$, of CdS NCs. But typically $\mu/e$ is a fraction of the particle radius.[23] By taking $n_0 = 2.4$ [24] and $\tau \approx 114$ ps [16] from published reports and the values of $\mu/e$, $T_2$, $\Delta N_0$ and $\lambda_0$ given in Table II, we apply Eqs. (2) and (4) to fit the experimental $\alpha_2$ data measured at different wavelengths by nanosecond Z-scans. Herein we assume that the relaxation time $\tau$ does not vary with particle size.[25] The third-order $\alpha_2$ wavelength dependences (stars and squares) and the corresponding two-level model fitting curves (solid lines) are plotted in Fig. 6(a) for Sample 1 and Sample 2. The theoretical and experimental data are in good agreement. Our $\mathrm{Im}\chi^{(3)}$ values qualitatively agree with the published results.[25, 26] Apparently, the value of $\mathrm{Im}\chi^{(3)}_{QD}$ is significantly increased as the investigated wavelength approaching the exciton peak. Although the $\alpha_2$ values are similar for the two samples, the intrinsic $\mathrm{Im}\chi^{(3)}_{QD}$ is greatly enhanced for the smaller CdS particle size by taking consideration of the different volume fraction in the two samples. It is consistent with the prediction from strong quantum confinement effect.[19, 27] By considering the volume fraction and the local field effect, the nonlinear absorption in CdS NCs is much higher than the value of bulk CdS.[28] The enhancement of the third-order nonlinearity can be attributed to the concentration of exciton oscillator strength.[29] As the particle size is reduced, a series of nearby transitions occurring at slightly different energies in the bulk are compressed by quantum confinement into a single, intense transition in a quantum dot. Therefore, the oscillator strength of the NC is concentrated into just a few transitions and the strong exciton bleaching can be expected. At room temperature, suppose that all the charged carriers stay in the ground state. So the population difference at zero field, $\Delta N_0$, is equal to the density of two-level atoms in the ensemble. From the point view of CdS NCs, the dipole moment $\mu$ should scale as the oscillator strength of the lowest excitonic transition. Herein we ignore the effects of biexciton and the interaction between exciton and exciton. Therefore, it is reasonable that the greater $\mu$, as illustrated in Table II, is obtained by fitting the experimental data of CdS NCs with the smaller size. According to the two-level model discussed above, the excitonic bleaching of the CdS NCs originates from the transition between the ground state 1S(e) and the lowest excited state $1S_{3/2}$(h).

We also apply Eqs. (3) and (5) to fit the experimental $\alpha_4$ data obtained at different wavelengths with the same physical parameters given in Table II. The wavelength dependences of $\alpha_4$ (stars and squares) and the corresponding two-level model fitting curves (solid lines) are plotted in Fig. 6(b) for Sample 1 and Sample 2. The theoretical curves have a similar trend to the experimental data. The discrepancy between the theoretical and experimental data is quite reasonable since we ignore the contribution from the higher excitonic transitions. In addition, the lack of accurate energy relaxation time $\tau$ may contribute to the discrepancy. At the high input irradiance of 57 MW/cm$^2$, shown in Fig. 4(b), the transmission drop at the focal plane is a result of fifth-order nonlinearity since $\alpha_4$ and $\alpha_2$ have opposite sign. When the laser intensity is

~35 MW/cm$^2$ (0.17 J/cm$^2$) or greater, the induced fifth-order nonlinearity is larger than the third-order nonlinear process, and hence the transmission drops. As expected, we also observe the linear dependence of both $\alpha_2$ and $\alpha_4$ values on the concentration of the CdS NCs. Our attempts to determine the nonlinear refraction were unsuccessful due to the presence of the high nonlinear absorption and the opposite contribution of the third-order and fifth-order nonlinearity. According to the absorption spectra shown in Fig. 3(a), the excitation at 420 and 430 nm corresponds to the absorption close to the resonance of the 1S(e) – $1S_{3/2}$(h) excitonic transition. Our Z-scan results at 420 and 430 nm reveal a transmission drop occurred at the higher irradiances, which is attributed to the saturation of the excitonic nonlinear absorption. In a two-level model, the absorption coefficient can be written as[30]

$$\alpha = \frac{\alpha_0}{\sqrt{1 + \mathrm{I}/\mathrm{I}_s}} \tag{6}$$

where $\alpha_0$ is the linear absorption coefficient, $I$ the laser irradiance and $I_s$ the saturation intensity. The absorption coefficient $\alpha$ can be expanded in terms of $I$ and $\alpha_2$ and $\alpha_4$ can be written as $\alpha_2 = -\alpha_0/2I_s$ and $\alpha_4 = 3\alpha_0/8I_s^2$, respectively. The obtained $I_s$ values at different wavelengths are listed in Table I. It is noted that the saturation intensity is nearly a constant at all the wavelengths close to the resonance of the 1S(e)-$1S_{3/2}$(h) exciton for both Sample 1 and Sample 2. Even though considering the correction of volume fraction, our saturation intensity in CdS NCs (1.2 ~ 1.5 MW/cm$^2$) is much lower in comparison to that found in bulk CdS ($I_s$ ~ 65 GW/cm$^2$).[31] The saturation intensity of 2.3-nm-sized CdS NCs is nearly two times greater than that of 3.0-nm-sized CdS NCs. The increase of saturation intensity with decrease of particle size is attributed to the concentration of the excitonic oscillator strength, which results from the quantum confinement effect.

## IV. CONCLUSIONS

In summary, the dispersion and irradiance dependence of excitonic nonlinear absorption in CdS NCs with 3.0 and 2.3 nm sizes have been experimentally investigated by using nanosecond Z-scan and theoretically studied by applying the two-level model consisting of the quantized sublevels 1S(e) and $1S_{3/2}$(h). The nonlinear absorption is greatly enhanced when the particle size is less than the bulk Bohr diameter due to the quantum confinement effect. In addition, we observe third-order excitonic bleaching at low irradiance and fifth-order excitonic nonlinear absorption at higher irradiance for both 3.0-nm-sized and 2.3-nm-sized CdS NCs. The intrinsic excitonic nonlinear absorption is explained successfully by the two-level model.

## ACKNOWLEDGMENTS

The work is supported by Research Grant No. R-144-000-084-112 from the National University of Singapore.

## References

[1]D. Cotter, M. G. Burt, and R. J. Manning, Phys. Rev. Lett. **68,** 1200 (1992).

[2]G. P. Banfi, V. Degiorgio, D. Fortusini, and H. M. Tan, Appl. Phys. Lett. **67,** 13 (1995).

[3]V. Klimov, P. Haring Bolivar, and H. Kurz, Phys. Rev. B **53,** 1463 (1996).

[4]R. E. Schwerzel, K. B. Spahr, J. P. Kurmer, V. E. Wood, and J. A. Jenkins, J. Phys. Chem. A **102,** 5622 (1998).

[5]M.Y. Han, W. Huang, C. H. Chew, L. M. Gan, X. J. Zhang, and W. Ji, J. Phys. Chem. B **102,** 1884 (1998).

[6]D. Matsuura, Y. Kanemitsu, T. Kushida, C. W. White, J. D. Budai, and A. Meldrum, Appl. Phys. Lett. **77,** 2289 (2000).

[7]K. S. Bindra and A. K. Kar, Appl. Phys. Lett. **79,** 3761 (2001).

[8]H. Du, G. Q. Xu, W. S. Chin, L. Huang, and W. Ji, Chem. Mater. **14,** 4473 (2002).

[9]P. Nandakumar, C. Vijayan, and Y. V. G. S. Murti, J. Appl. Phys. **91,** 1509 (2002).

[10]Y. Wang and W. Mahler, Opt. Commun**. 61,** 233 (1987).

[11]E. F. Hilinski, P. A. Lucas, and Y. Wang, J. Chem. Phys. **89,** 3435 (1988).

[12]Y. Wang, A. Suna, J. McHugh, E. F. Hilinski, P. A. Lucas, and R. D. Johnson, J. Chem. Phys. **92,** 6927 (1990).

[13]H. Yao, S. Takahara, H. Mizuma, T. Kozeki, and T. Hayashi, Jpn. J. Appl. Phys. **35,** 4633 (1996).

[14]T. Miyoshi, N. Matsuo, P. Maly, F. Trojanek, P. Nemec, and J. Kudrna, J. Mater. Sci. Lett. **20,** 343 (2001).

[15]V. I. Klimov, Ch. J. Schwarz, and D. W. McBranch, Phys. Rev. B **60,** R2177 (1999).

[16]G. H. Ma, S. H. Tang, W. X. Sun, Z. X. Shen, W. M. Huang, and J. L. Shi, Phys. Lett. A **299,** 581 (2002).

[17]Z. H. Zhang and W. S. Chin, submitted to Chem. Mater.

[18]B. D. Cullity, Elements of X-ray Diffraction (Addison-Wesley, New York, 1977).

[19]Y. Wang and N. Herron, Phys. Rev. B **42,** 7253 (1990).

[20]J. B. Li and J. B. Xia, Phys. Rev. B **62,** 12613 (2000).

[21]M. Sheik-Bahae, A. A. Said, T. H. Wei, D. J. Hagan, and E. W. Van Stryland, IEEE J. Quantum Electron. **26,** 760 (1990).

[22]A Yariv, Quantum Electronics (John Wiley & Sons, New York, 1975), page 153.

[23]M. E. Schmidt, S. A. Blanton, M. A Hines, and P. Guyot-Sionnest, Phys. Rev. B **53,** 12629 (1996).

[24]Semiconductor: Physics of II-VI and I-VII Compounds and Semimagnetic Semiconductors, edited by O. Madelung, M. Schultz, and H. Weiss, Landolt-Bonstein, New Series, Group III, Vol. **17**. Subvol. B (Sringer, Berlin, 1982).

[25]P. Nandakumar, C. Vijayan, and Y. V. G. S. Murti, Opt. Commun. **185,** 457 (2000).

[26]T. Takada, J. D. Mackenzie, M. Yamane, K. Kang, N. Peyghambarian, R. J. Reeves, E. T. Knobbe, and R. C. Powell, J. Mater. Sci. **31,** 423 (1996).

[27]Y. Wang and N. Herron, J. Phys. Chem. **95,** 525 (1991).

[28]H. P. Li, C. H. Kam, Y. L. Lam, and W. Ji, Opt. Commun. **190,** 351 (2001).

[29]A. P. Alivisatos, Science **271,** 933 (1996).

[30]R. L. Sutherland, Handbook of Nonlinear Optics (Marcel Dekker, Inc., 1996).

[31]J. F. Lami, P. Gilliot, and C. Hirlimann, Phys. Rev. Lett. **77,** 1632 (1996).

Figure Captions:

FIG. 1 (a) TEM image of CdS nanocrystals capped with dodecanethiol (Sample 2). (b) Size distribution of CdS nanocrystals. The solid line is a lognormal fit.

FIG. 2 The powder X-ray diffraction pattern (······) of CdS nanocrystals with 2.3-nm diameter (Sample 2) along with Gaussian fit (—). The deconvoluted individual reflection peaks are indicated by dashed lines.

FIG. 3 Optical absorption spectra of CdS nanocrystals fitted to three Gaussian bands: (a) Sample 1; (b) Sample 2.

FIG. 4 (a) Open aperture Z-scans measured at different wavelengths for 3.0-nm-sized CdS nanocrystals ($4 \times 10^{-3}$ Mol/L) with 5-ns OPO laser pulses. The input irradiances are 0.012 GW/cm$^2$. The scatter graphs are experimental data while the solid lines are theoretically fitting curves by employing the standard Z-scan theory. For clear presentation, the curves are vertically shifted by 0.4, 0.3, 0.2, 0.1 and 0.0, respectively.

(b) Comparison of open aperture Z-scan measurements at different irradiances at 430 nm for 3.0-nm-sized CdS nanocrystals ($2 \times 10^{-3}$ Mol/L) by using 5-ns OPO laser pulses. The scatter graphs are experimental data while the solid lines are theoretically fitting curves by employing the standard Z-scan theory. For clear presentation, the graphs are vertically shifted by 0.2, 0.1 and 0.0, respectively.

(c) The effective nonlinear absorption coefficient $\alpha_{2eff}$ of 3.0-nm-sized CdS nanocrystals ($4 \times 10^{-3}$ Mol/L) plotted as a function of the input irradiance. The solid line is the linear fit of the data.

FIG. 5 Wavelength dependence of open aperture Z-scan measurements at high irradiance for 3.0-nm-sized CdS nanocrystals ($2 \times 10^{-3}$ Mol/L) by using 5-ns OPO laser pulses. The scatter graphs are experimental data while the solid lines are theoretically fitting curves by employing the standard Z-scan theory. For clear presentation, the graphs are vertically shifted by 0.4, 0.3, 0.2, 0.1 and 0.0, respectively.

FIG. 6 The third-order nonlinear absorption coefficient $\alpha_2$ (a) and fifth-order nonlinear absorption coefficient $\alpha_4$ (b) of 3.0-nm-sized (Sample 1, stars) and 2.3-nm-sized (Sample 2, squares) CdS nanocrystals plotted as a function of wavelength. The solid curves are the theoretically fitting results by employing the two-level model.

TABLE I. Nonlinear absorption $\alpha_2$, $Im\chi^{(3)}_{QD}$, $\alpha_4$, $Im\chi^{(5)}_{QD}$ and saturation intensity $I_s$ at different wavelengths for Sample 1 and Sample 2.

| | Sample 1 | | | | | Sample 2 | | | | |
|---|---|---|---|---|---|---|---|---|---|---|
| λ (nm) | $\alpha_2$ (cm/GW) | $Im\chi^{(3)}_{QD}$ ($10^{-6}$esu) | $\alpha_4$ ($10^3 cm^3/GW^2$) | $Im\chi^{(5)}_{QD}$ ($10^{-11}$esu) | $I_s$ ($MW/cm^2$) | $\alpha_2$ (cm/GW) | $Im\chi^{(3)}_{QD}$ ($10^{-6}$esu) | $\alpha_4$ ($10^3 cm^3/GW^2$) | $Im\chi^{(5)}_{QD}$ ($10^{-11}$esu) | $I_s$ ($MW/cm^2$) |
| 430 | -299 | -2.29 | 8.3 | 3.5 | 1.4 | -328 | -5.01 | 12.4 | 10.4 | 1.3 |
| 435 | -218 | -1.68 | 5.6 | 2.4 | 1.4 | -282 | -4.36 | 9.6 | 8.1 | 1.2 |
| 440 | -152 | -1.19 | 4.5 | 1.9 | 1.5 | -179 | -2.81 | 5.1 | 4.4 | 1.4 |
| 445 | -148 | -1.17 | 3.5 | 1.5 | 1.3 | -146 | -2.32 | 4.8 | 4.1 | 1.3 |
| 450 | -100 | -0.806 | 3.3 | 1.4 | 1.3 | -116 | -1.87 | 4.3 | 3.7 | 1.2 |

TABLE II. The physical parameters used in the two-level model.

| | Size (nm) | $\lambda_0$ (nm) | $\alpha_0$ ($cm^{-1}$) | $\Delta N_0$ ($10^{15}/cm^3$) | $\tau$ (ps) | $T_2$ (fs) | $\mu/e$ (nm) |
|---|---|---|---|---|---|---|---|
| Sample1 | 3.0 | 400 | 2.12 | 8.5 | 114 | 2.9 | 0.33 |
| Sample2 | 2.3 | 380 | 2.87 | 9.4 | 114 | 2.9 | 0.45 |

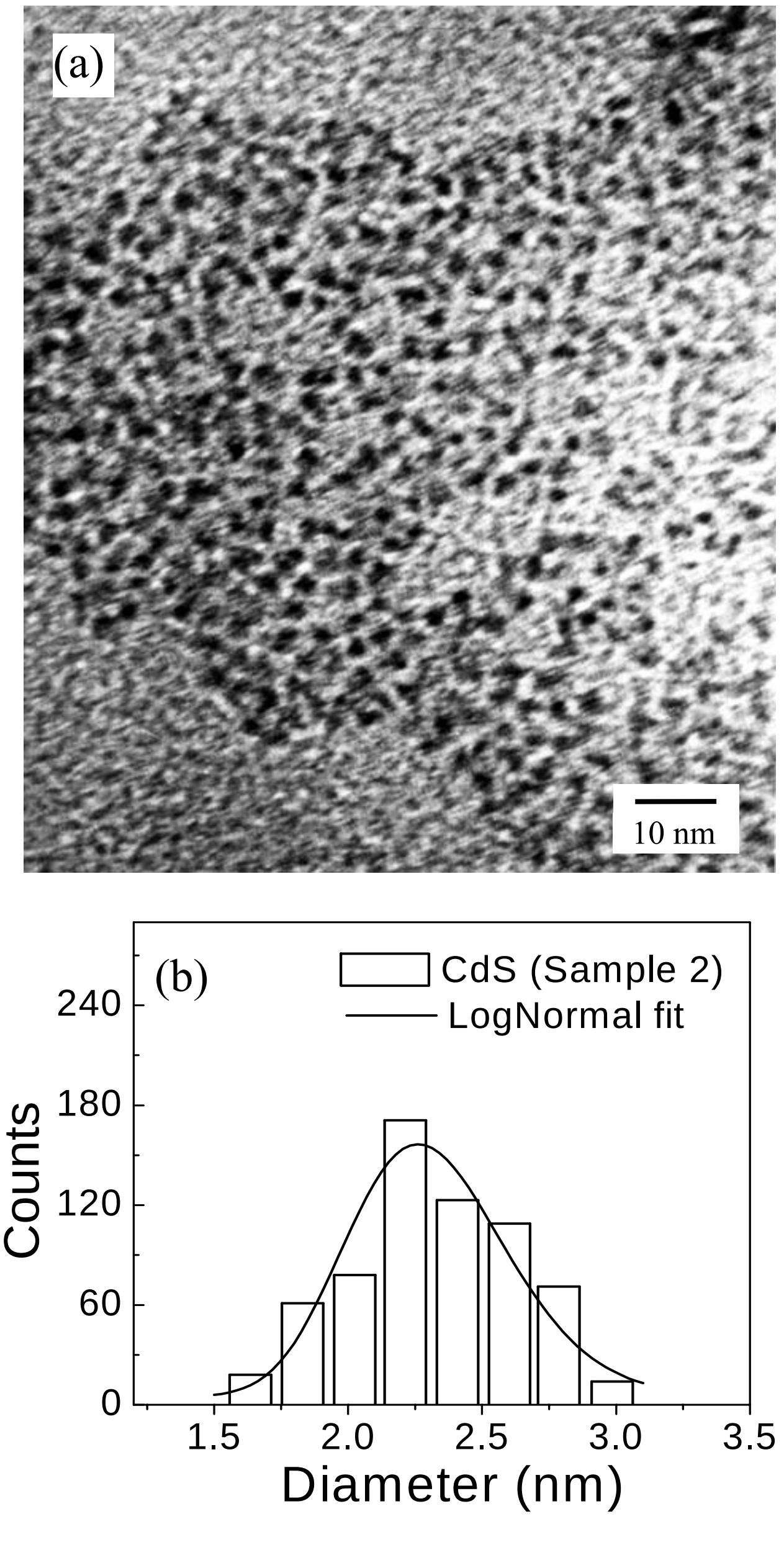

**Fig.1 J. He et al.**

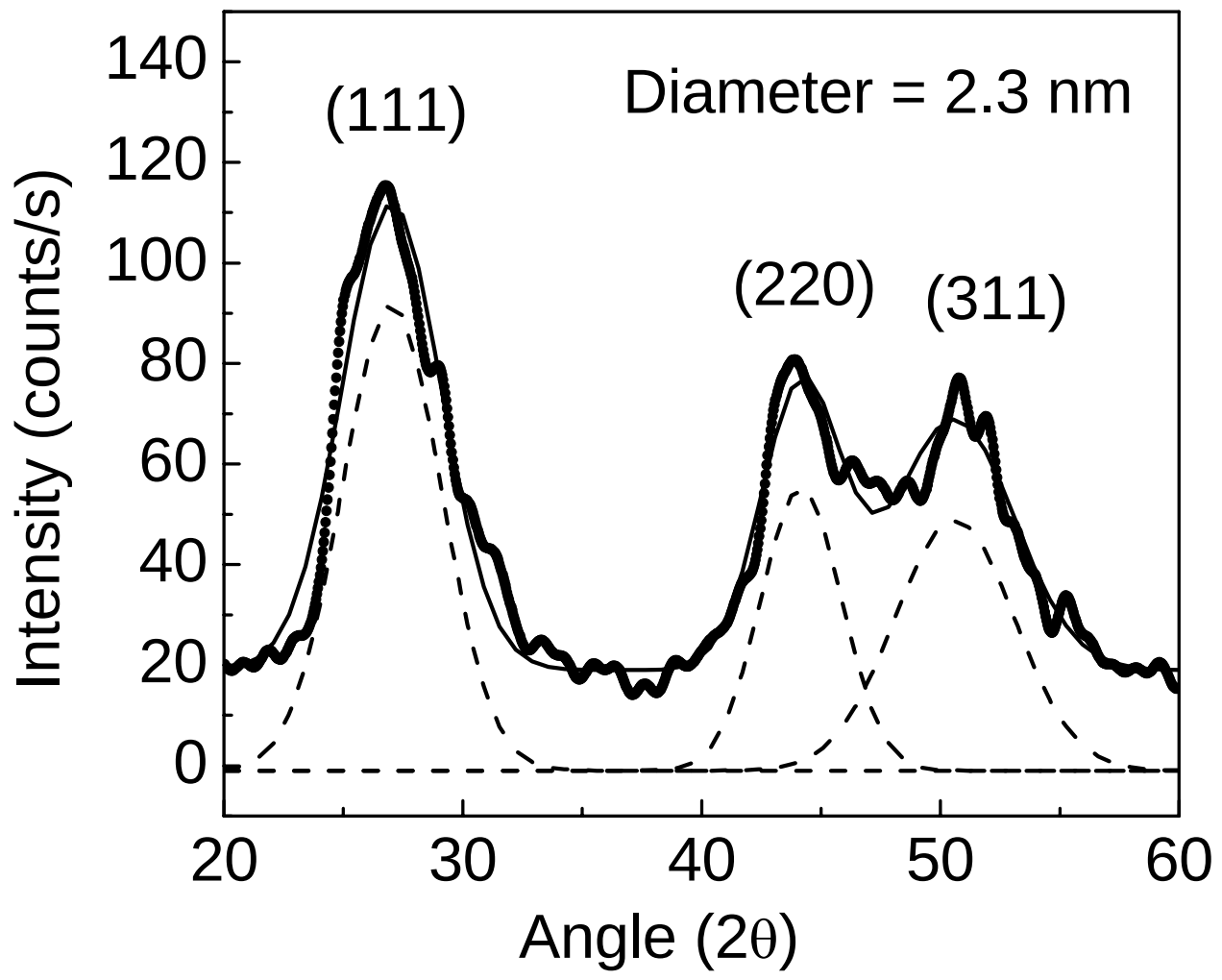

**Fig.2 J. He et al.**

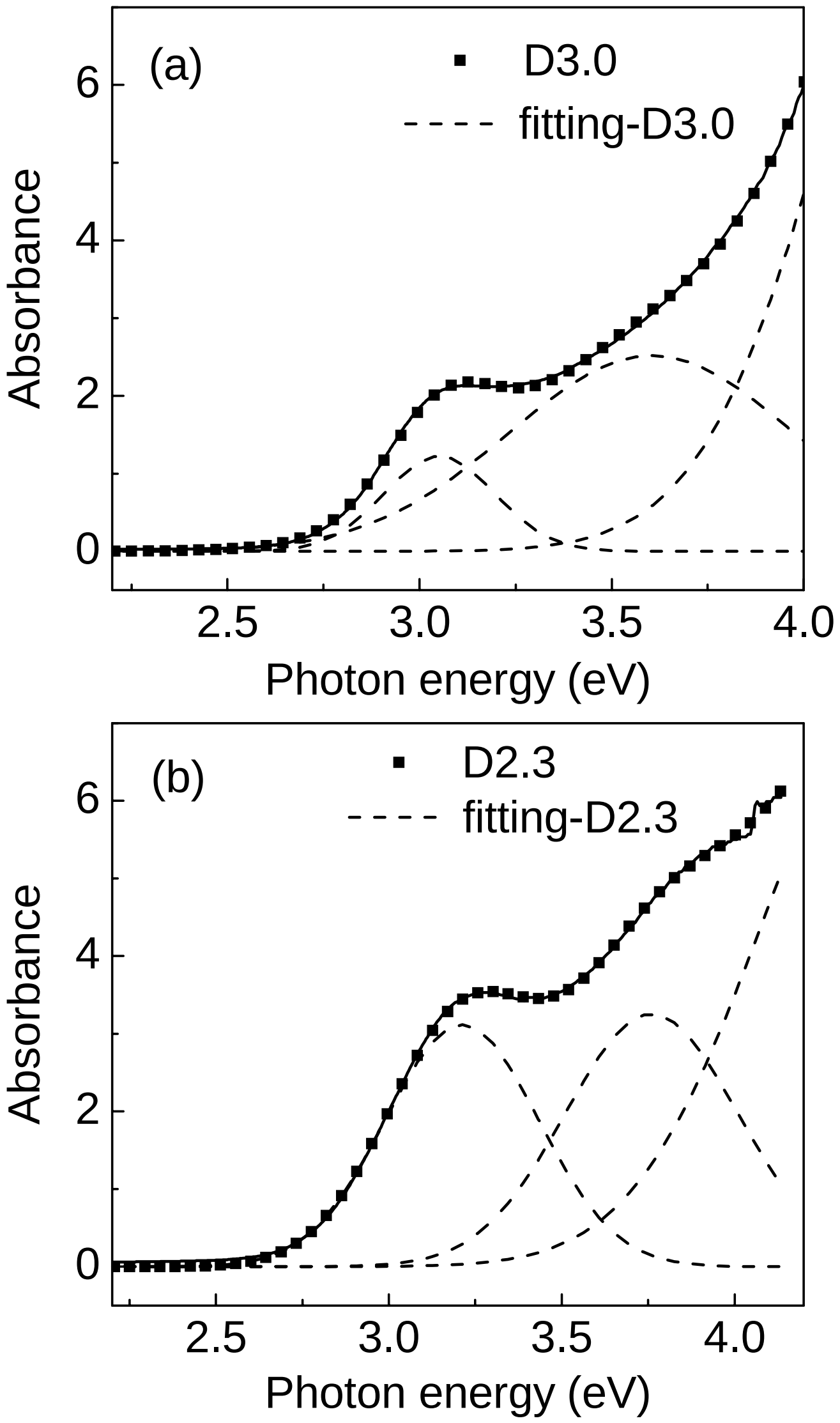

**Fig. 3 J. He et al.**

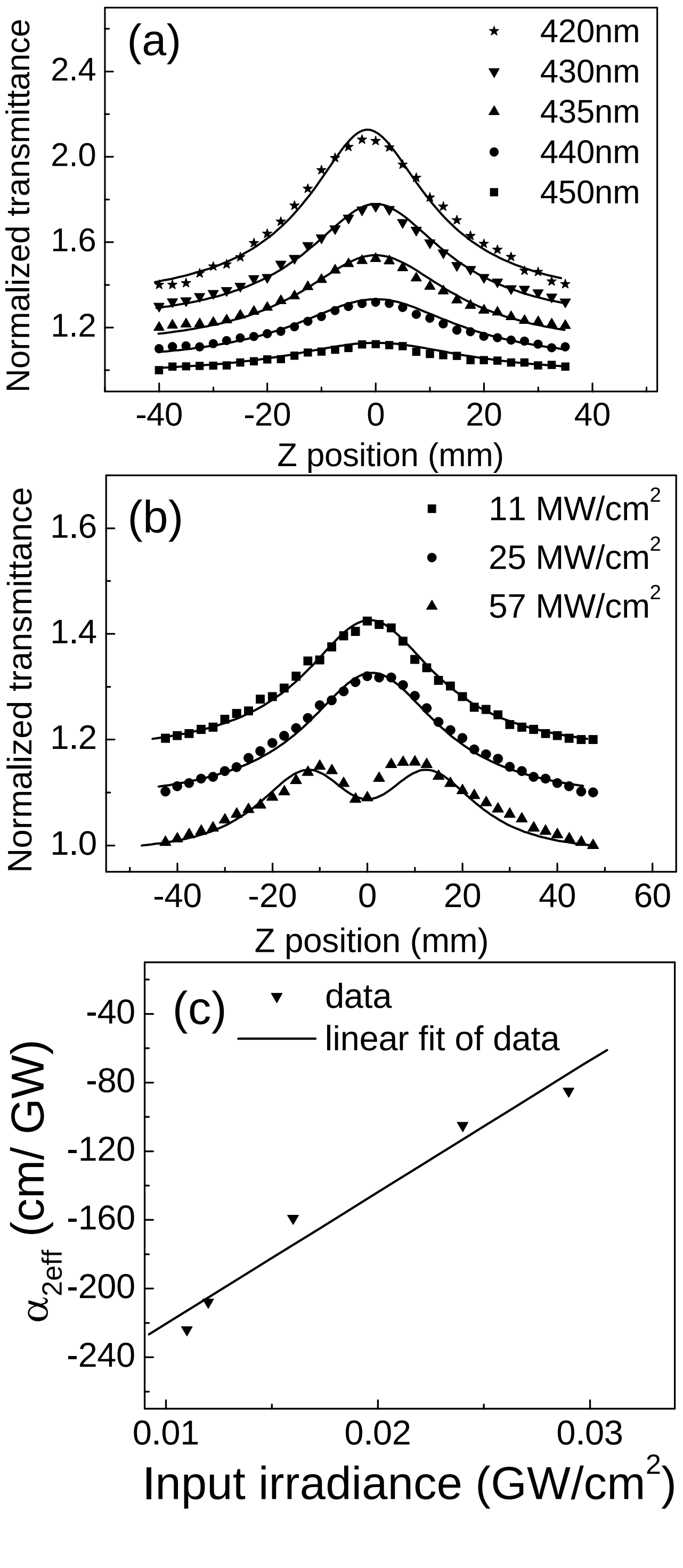

**Fig. 4 J. He et al.**

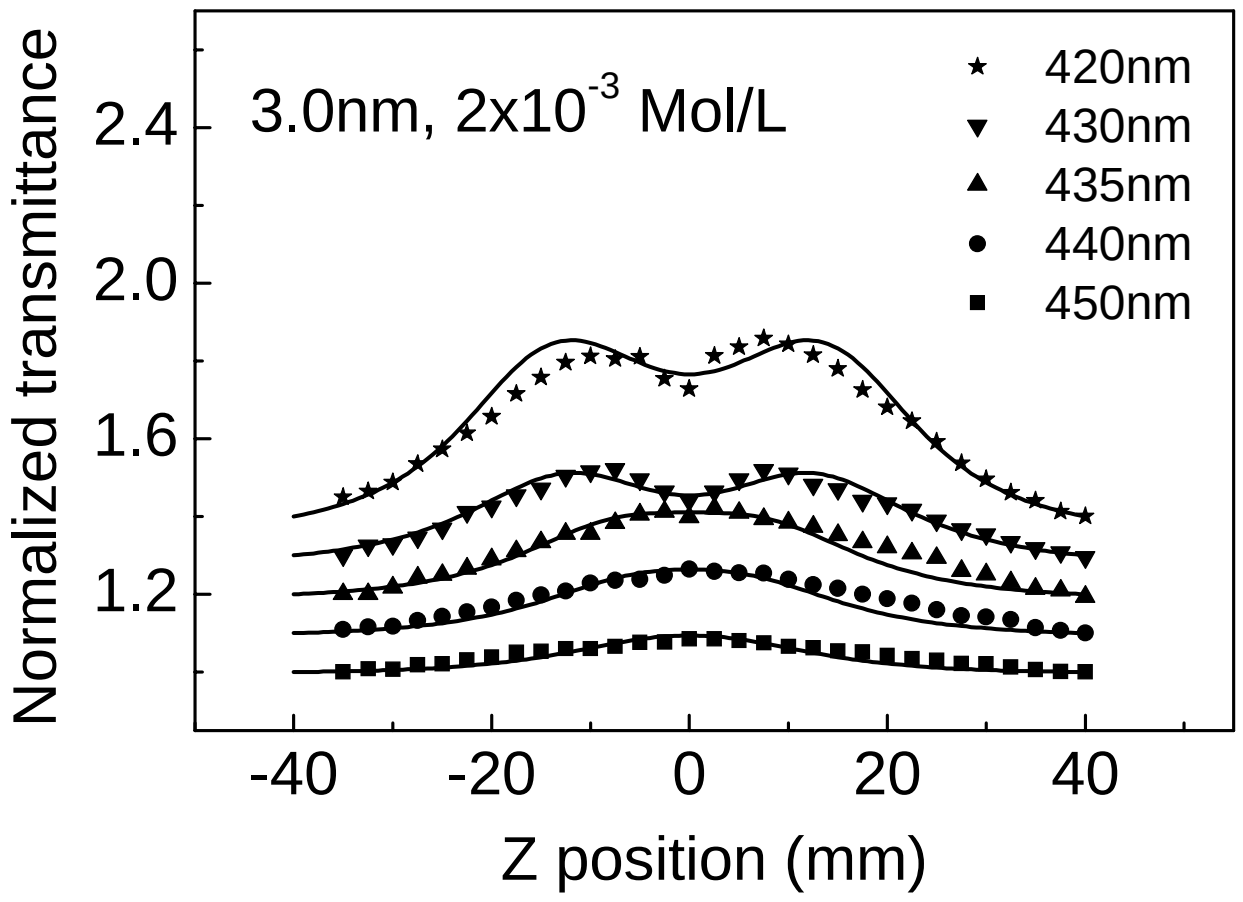

**Fig. 5 J. He et al.**

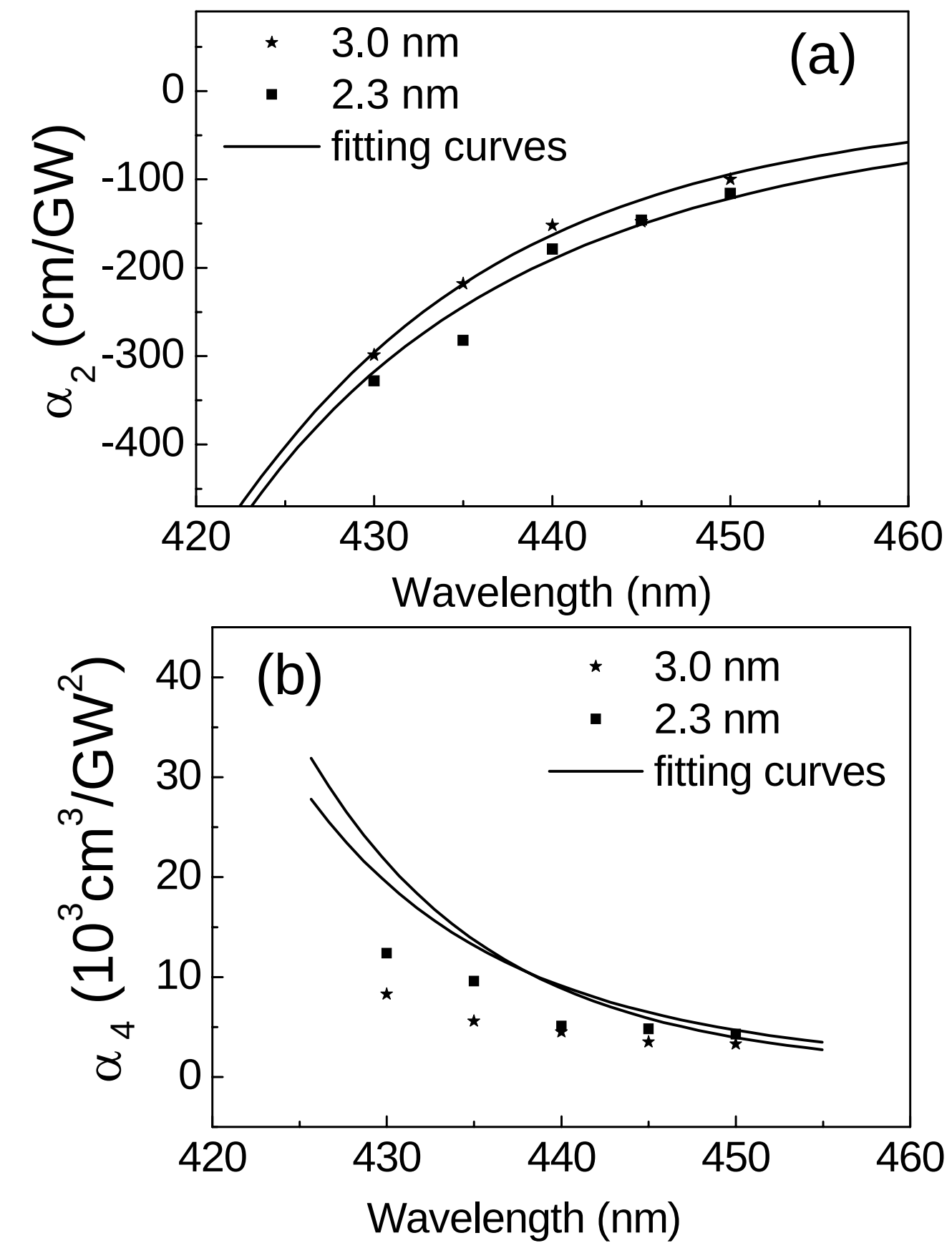

**Fig. 6 J. He et al.**